\documentclass[11pt]{article}
\usepackage[centertags]{amsmath}
\usepackage{amsfonts} \usepackage{amssymb} \usepackage{amsthm}

\def\one{{\rm 1\kern -.9mm l}}                             %

\def\beq{\begin{equation}}
\def\eeq{\end{equation}}
\def\beqa{\begin{eqnarray}}
\def\eeqa{\end{eqnarray}}
\begin{document}

\begin{titlepage}
 \vspace*{ 3.0cm}
\centerline{\LARGE \bf Anomalies and Tadpoles }
\vskip .5cm
\centerline{\LARGE \bf in} \vskip .5cm
\centerline{\LARGE \bf Open/Closed String Duality }
\vskip .8cm
\centerline{\bf 
A. Liccardo, R. Marotta and F. Pezzella}
\vskip .4cm \centerline{\sl Dipartimento di
Scienze Fisiche, Universit\`a di Napoli and INFN, Sezione di
Napoli} \centerline{\sl Complesso Universitario Monte
S. Angelo, ed. G  - via Cintia -  I-80126 Napoli, Italy}
 \vskip 2cm
\begin{abstract}
We discuss the role played by the divergences appearing
in the interaction between a fractional
D3 brane dressed with an $SU(N)$ gauge field and a stack of $N$ fractional
D3 branes on the orbifolds $C^2/Z_2$ and
$C^3/(Z_2 \times Z_2)$. In particular we show that the logarithmic
divergences in the closed string channel, interpreted
as due to twisted massless tadpoles, are mapped, under open/closed
string duality, in the logarithmic ones in the
open string channel, due to the massless states circulating in the
annulus diagram and corresponding to the one-loop divergences that one 
finds in the gauge theory
living in the world volume of the brane.
This result provides a quantitative evidence of why the chiral
and scale anomalies of the supersymmetric and non conformal gauge
theories supported by the world volume of the
branes can be inferred from supergravity calculations.
\end{abstract}
\vfill  {\small Work partially supported by the European
Commission RTN Programme HPRN-CT-2000-00131 and by MIUR.}
\end{titlepage}

\newpage

\section{Introduction}
\label{sez0}
This work, based on Ref. \cite{DLMP}, is aimed to a deeper understanding of 
the relation between the 
{\em gauge/gravity correspondence} and the
{\em open/closed string duality}. The former indicates the two-fold property 
exhibited by a D-brane of being a solution of the 
low-energy string effective action ({\em supergravity}) and of having open strings 
with their endpoints attached to its world-volume; the 
latter
denotes the equivalence between the open string annulus diagram, parametrized 
by the proper time $\tau$, and the closed string 
tree diagram, when the modular transformation $\tau \rightarrow 1/\tau$ 
is performed - for recent reviews on this subject, see for example
Ref. \cite{DL03}.

By using the gauge/gravity correspondence it has been 
shown 
that the classical supergravity solutions corresponding to fractional
D branes, stuck at the orbifold fixed points, encode perturbative properties 
of non conformal
gauge theories, with reduced supersymmetry, living on their world volume, 
such as
the chiral and scale anomaly\footnote{For a review of fractional branes
and their application to the study of non conformal gauge theories
see Refs.~\cite{REV} and \cite{MA}.}. Relevant improvements have been performed in 
deriving from such a kind of branes also non-perturbative properties of the gauge theories \cite{MS} \cite{IL}. Since, as it is well-known, gauge theories and 
supergravity are related to open and closed strings
respectively, it is quite natural to investigate on the possibility of 
deriving the gauge theory anomalies by only using the 
open/closed string duality in a pure stringy framework. In other words, one 
can ask whether the gauge/gravity correspondence could be 
seen as a direct consequence of the open/closed string duality. 
The answer we have got is positive, at least in the cases we have considered. 
More precisely, we can state that the quantum ultraviolet properties of 
the gauge theory living on a D-brane can be equivalently 
derived by performing the field theory limit either in the open 
string channel, as expected, or in the closed string channel and in 
this case
the gauge theories anomalies can be inferred from logarithmic 
divergences due to the exchange of twisted massless states 
({\em twisted tadpoles}). 
Hence a clear and beautiful relation between the gauge/gravity 
correspondence and the open/closed string duality, lying in turn on 
the relation between the gauge theory anomalies and the presence of 
twisted massless states in the closed string channel, 
emerges out. This has been the main goal of our work. 

Our starting point has been the computation of the one-loop string annulus 
diagram which describes the interaction of a fractional brane of the orbifold
$C^2/Z_2$ having an $SU(N)$ gauge field on it with $N$ fractional branes 
without any gauge field and we have extracted from it the coefficient of the gauge
kinetic term. This contribution results to be logarithmically divergent
at the string level \cite{LR} \cite{KAKURO} \cite{NAPOLI}. 
But, differently from other kinds of 
higher divergences which are in general a sign of inconsistency of the theory and therefore have 
to be eliminated in some way (see for instance Refs. \cite{FS} and 
\cite{CLNY}), this 
logarithmic divergence can be interpreted, 
when read from the point of view of the closed string channel, as due to the
exchange of twisted massless tadpoles.
The logarithmic nature of this divergence derives from the fact that 
such an exchange occurs in 
the two directions transverse 
to the branes and the orbifold. Furthermore this divergence is
ultraviolet in the open string channel and infrared in the closed string one, 
therefore we have regularized the string calculation by introducing an ultraviolet cutoff in the former corresponding to an infrared cutoff in the latter.
We have shown that, under the modular transformation that maps the open string channel into the closed string one, {\em open string massless states go into closed string massless states, and open string massive states go into closed string massive states, without any mixing between massless and massive states!}
Hence the divergent contribution can be seen to come from the exchange of 
massless closed string states between the two fractional
branes or from the massless open string states circulating in the loop.
By adding also the contribution of 
the open string tree diagrams to the one-loop string diagram, we find an expression for the
gauge coupling constant that gives the correct beta-functions of ${\cal N}=2$ and ${\cal N}=1$.
This result provides a quantitative evidence of why the one loop-beta 
function can be derived from supergravity calculations \cite{d3} \cite{polch} \cite{marco} 
\cite{grana2} \cite{d3d7}\cite{FERRO}\cite{ANOMA}.

\section{Branes in External Fields: ${\cal N}$ = 2 orbifold}
\label{sez1}

We analyze the interaction between a D$3$ brane with an external
$SU(N)$ gauge field on its world-volume and a stack of $N$
ordinary D$3$ branes. This can be equivalently done by computing either, in the
open string channel, the one-loop open string diagram
or, in the closed string channel, the tree closed
string diagram containing two boundary states and a closed string
propagator.

In order to obtain a gauge theory with reduced supersymmetry we
consider Type IIB superstring theory on an orbifold space.
Furthermore, to get a non conformal gauge theory, we study
fractional D3 branes which are characterized by their
being stuck at the orbifold fixed point and which, unlike bulk branes,
have a non conformal theory on their world-volume. For the sake of
simplicity we consider fractional branes of the 
orbifold ${\rm I\! R}^{1,5} \times
 C^2 / Z_2$ supporting ${\cal{N}}=2$ super Yang-Mills  on their
world volume.

The $ Z_2$ group, that is chosen to be acting on the coordinates
$x^{m}$ with
$m=6,7,8,9$, is characterized by two elements $(e,h)$, being $e$
the identity element and $h$ such that $h^2=e$. The element
$h$ acts on the complex combinations
$\vec{z} = (z^1, z^2)$, where $z^1=x^6+ix^7$,
$z^2=x^8+ix^9$, as
$ (z_1 \,,\,z_2)\rightarrow ( - z_1 \,,\, - z_2)$.
The orbifold group $Z_2$ acts also on the Chan-Paton factors
located at the endpoints of the open string stretched between
the branes.  Fractional
branes are
defined as branes for which such factors transform according to irreducible
representations of the orbifold group and
we
consider only the trivial one corresponding to a particular kind of
fractional branes. The orbifold we are
considering is non compact and has therefore
only one fixed point located at $z_1=z_2=0$.
We are interested in the case of parallel fractional D3 branes with
their world volume along the directions $x^0 , x^1 , x^2, x^3$, that are
completely external to the space on which the orbifold acts.
The gauge field lives on the four-dimensional world volume of the
fractional D3 brane and can be chosen to have the following non
vanishing entries:
\begin{eqnarray}
\hat{F}_{0 1 }= -\hat{F}_{ 10 }=f \qquad \hat{F}_{23}= -\hat{F}_{32}=g ,
\label{effe}
\end{eqnarray}
where $\hat F_{\alpha\beta}=2\pi\alpha'F_{\alpha\beta}$.

The interaction between two branes is given by the vacuum
fluctuation of an open string that is stretched between them. In
particular, the free energy of an open string stretching between a
D3 brane and a stack of $N$ D3 branes located at a distance $y$ in
the plane ($x^4 , x^5$), orthogonal to both the D3 brane
world-volume and the four-dimensional space on which the orbifold
acts, is given by \cite{DLMP}:
\begin{eqnarray}
&&\hspace{-2cm}Z_h^o  = \frac{N}{(8 \pi^2 \alpha')^2} \int d^4x
\sqrt{ -\mbox{det}(\eta+\hat{F})}
\int_{0}^\infty \frac{d\tau}{\tau}e^{-\frac{y^2 \tau}{2\pi\alpha'}}
\frac{ 4\, \sin \pi \nu_f  \sin \pi \nu_g}
{\Theta_{2}^2(0|i\tau) \Theta_{1}(i\nu_f\tau|i\tau)\Theta_{1}
(i\nu_g\tau|i\tau)}
\nonumber \\
{} &&\hspace{-1.5cm} \times \left[ \Theta_{3}^2(0|i\tau)\Theta_{4}(i\nu_f\tau|i\tau)
\Theta_{4}(i\nu_g\tau|i\tau) -
\Theta_{4}^2(0|i\tau) \Theta_{3}(i\nu_f\tau|i\tau)
\Theta_{3}(i\nu_g\tau|i\tau) \right] \nonumber \\
&& \hspace{-1.5cm} + \frac{iN}{32\pi^2}\int d^4x\, F_{\alpha\beta}^a
{\tilde F}^{a\,\alpha\beta}
\int_{0}^\infty \frac{d\tau}{\tau}e^{-\frac{y^2 \tau}{2\pi\alpha'}},
\label{zetatet}
\end{eqnarray}
where $\tilde{F}_{\alpha \beta }=\frac{1}{2}\epsilon _{\alpha \beta \gamma
\delta }F^{\gamma \delta }$ and
\begin{eqnarray}
\nu_{f} \equiv \frac{1}{2 \pi i}\log \frac{1+f}{1-f}
\,\,\,\,\,\,\,\,\,\,\,{\rm and}\,\,\,\,\,\,\,\,\,\,\,
\nu_{g} \equiv \frac{1}{2 \pi i}\log \frac{1-i g}{1+i g} .
\label{nufnug2}
\end{eqnarray}
Here  we have only considered the term in the free energy expression
coming from the non trivial element $h$ of the orbifold group.
Indeed this is the only one that contributes to the one-loop
anomalies of the gauge theory in the world volume, as we will see later.

The above computation can also be  performed in the
{\em closed string channel} where the relevant contribution is due to the
tree level propagation of twisted states, that is:
\begin{eqnarray}
&&\hspace{-2cm}Z_h^c =\frac{N}{(8\pi^2\alpha')^2} \int d^4x
\sqrt{ -\mbox{det}(\eta+\hat{F})}
\int_{0}^\infty \frac{dt}{t}e^{-\frac{y^{2} }{2\pi\alpha't}}
\frac{4\,\sin\pi\nu_f \sin\pi\nu_g}{
\Theta_{4}^2(0|it) \Theta_{1}(\nu_f|it)\Theta_{1}(\nu_g|it)}
\nonumber\\
&&\hspace{-1.5cm}\times \left\{\Theta_{2}^2(0|it) \Theta_{3}(\nu_f|it)\Theta_{3}(\nu_g|it)
-\Theta_{3}^2(0|it) \Theta_{2}(\nu_f|it)\Theta_{2}(\nu_g|it) \right\}
\nonumber\\
&&\hspace{-1.5cm}+\frac{iN}{32\pi^2} \int d^4x
F^a_{\alpha\beta}\tilde F^{a\alpha\beta}\int \frac{dt}{t}
e^{-\frac{y^{2} }{2\pi\alpha't}}.
\label{closed}
\end{eqnarray}
The two expressions for $Z$ separately obtained in the open and
the closed string channels are, as expected, equal to each other.
This equality is the essence of the open/closed string duality
and can be explicitly shown by using the modular
transformation that relates the modular parameters in the open and
closed string channels, namely $\tau=1/t$.
It is also easy to see that the distance $y$ between the dressed
D3 brane and the stack of the $N$ D3 branes  makes the integral in
Eq.~(\ref{closed}) convergent for small values of $t$, while in
the limit $t \rightarrow \infty$, the integral is logarithmically
divergent. This divergence is due to a twisted tadpole corresponding
to the exchange of massless closed string states between the two
stucks of branes. We would like to stress that
the presence of the gauge field
makes the divergence to appear already {\em at the string level}, before
any field theory limit ($\alpha ' \rightarrow 0$) is performed.
When $F$ vanishes, the divergence is eliminated by the integrand
being identically zero as a consequence of the fact that
fractional branes are BPS states.

Tadpole divergences
correspond in general to the presence of gauge anomalies, which
make the gauge theory inconsistent and must be eliminated by
drastically modifying the theory or by fixing particular values of
the parameters. Instead, as stressed in Refs.~\cite{LR}~\cite{KAKURO}~\cite{BM0002}~\cite{BIANCHI},
logarithmic tadpole divergences do not correspond to gauge
anomalies. In our case they correspond to the fact that the gauge
theory living on the brane is not conformally invariant. In fact,
they provide the correct one-loop running coupling constant. We
cure these divergences just by introducing in Eq. (\ref{closed})
an infrared cutoff $\Lambda$ that regularizes the contribution of the
massless closed string states. Since, in the open/closed string
duality, an infrared divergence in the closed string channel
corresponds to an ultraviolet divergence in the open string
channel, it is easy to see that the expression in Eq.
(\ref{zetatet}) is divergent for small values of $\tau$ and needs
an ultraviolet cutoff. This divergence is exactly the one-loop
divergence that one gets in ${\cal{N}}=2$ super Yang-Mills, which
is the gauge theory living in the world volume of the fractional
D3 brane.

From Eqs.~(\ref{zetatet}) and ~(\ref{closed}), which contain
arbitrary powers of the gauge field $F$, we extract  the quadratic
term in $F$. This term in the open string channel is given by:
\begin{eqnarray}
Z_h^o(F)\!\!&\rightarrow&\!\!\left[- \frac{1}{4} \int d^4 x F_{\alpha
\beta}^{a} F^{a\,\alpha \beta } \right] \nonumber \\
& \times &\!\!
\left\{
\frac{1}{g_{YM}^{2} (\Lambda )}
 - \frac{N}{8 \pi^2}
\int_{ {\small \frac{1}{\alpha' \Lambda^2}} }^{\infty} \frac{d \tau}{\tau}
{e}^{-\frac{y^{2} \tau}{2 \pi \alpha' } } + \frac{N}{8 \pi^2}
\int_{0}^{\infty} \frac{d \tau}{\tau}
{e}^{-\frac{y^{2} \tau}{2 \pi \alpha' } } G (k) \right\} \nonumber\\
&+&  iN \left[\frac{1}{32\pi^2} \int d^4x F^a_{\alpha\beta}\tilde
F^{a\,\alpha\beta} \right] \int_{  {\small \frac{1}{\alpha'
    \Lambda^2}} }^{\infty}
\frac{d\tau}{\tau}e^{-\frac{y^{2} \tau}{2\pi\alpha'}}
\label{gau57}
\end{eqnarray}
where we have  isolated the massless open string states, which are the
only ones giving a divergence, from the massive states encoded
in the function $G(k)$ so defined:
\begin{eqnarray}
G (k) = -\left[\frac{f_3(k)f_4(k)}{f_1(k)f_2(k)}\right]^4
2k\frac{d}{dk} \log\left[\frac{f_3(k)}{f_4(k)}\right] +1,
\label{Gk89}
\end{eqnarray}
Analogously in the closed
channel one gets:
\begin{eqnarray}
Z_h^c(F)\!\!&\rightarrow&\!\!\left[- \frac{1}{4} \int d^4 x
F_{\alpha \beta}^{a}
F^{a\,\alpha \beta }\right] \nonumber \\
& \times & \! \!
 \left\{
\frac{1}{g_{YM}^{2}(\Lambda ) }
- \frac{N}{8 \pi^2}
\int_{0}^{\alpha' \Lambda^2}
\frac{dt}{t}  {e}^{- \frac{y^{2}}{2 \pi\alpha' t} }
+\frac{N}{8 \pi^2}
\int_{0}^{\infty} \frac{dt}{t}  {e}^{- \frac{y^{2}}{2 \pi\alpha' t} }
F(q) \right\}  \nonumber\\
&+&  iN \left[ \frac{1}{32\pi^2} \int d^4x
F^a_{\alpha\beta}\tilde F^{a\,\alpha\beta} \right]
\int_{0}^{ {\alpha' \Lambda^2c }} \frac{dt}{t}
e^{-\frac{y^{2} }{2\pi\alpha't}}
\label{F285}
\end{eqnarray}
where, again, the massive states contribute as:
\begin{eqnarray}
F(q)= \left[\frac{f_3(q)f_2(q)}{f_1(q)f_4(q)}\right]^4
2q\frac{d}{dq} \log\left[\frac{f_3(q)}{f_2(q)}\right]  + 1~~.
\label{Fq89}
\end{eqnarray}
It turns out that the divergence that we have at the string level
is exactly the one-loop divergence present in the gauge field
theory living in the world volume of the brane and can be
equivalently seen as due, in the open channel, to the massless
open string states circulating in the loop and, in the closed
channel, to the massless closed string states exchanged between
two branes.

Notice that in the two previous equations we have also added the
contribution coming from the tree diagrams that contain the bare
coupling constant. In an ultraviolet finite theory such as string theory we
should not deal with a bare and a renormalized coupling. On the other
hand, we have already discussed the fact that the introduction of a
gauge field produces a string amplitude that is divergent already at the
string level and that therefore must be regularized with the
introduction of a cutoff.

One can
see that the contribution of the massless states and the one of the
massive states transform respectively into each other without any mixing
between massless and massive states, as follows from the identity:
\begin{eqnarray}
F (q) = G (k)
\label{FqGk}
\end{eqnarray}
that can be easily proven using the modular transformations of the
functions $f_i$.

This means that the open/closed string duality exactly maps the
ultraviolet divergent contribution coming from the massless open
string states circulating in the loop - and that reproduces the
divergences of ${\cal{N}}=2$ super Yang-Mills living in the
world volume of the fractional D3 branes - into the infrared
divergent contribution due to the massless closed string states
propagating between the two branes. This leads to the first
evidence of why the one-loop running coupling constant can be
consistently derived from a supergravity calculation.

To show more explicitly the connections between open/closed
duality and gauge/gravity correspondence we perform the field
theory limit of the previous expressions. This means to 
perform the zero slope limit ($\alpha'\rightarrow 0$) {\em
together} with the limit in which the modular variables $t$ and
$\tau$ go to infinity, keeping fixed the dimensional Schwinger
proper times  $\sigma
 = \alpha' \tau$
and  $s = \alpha' t$  of the open and closed string, respectively. Indeed,
in these two limits the massive states contributions in
Eqs.~(\ref{gau57}) and~(\ref{F285}) - depending on $G(k)$ and
$F(q)$ respectively - vanish leaving only the divergences due to
the massless states.
By extracting the coefficient of the term $F^2$ from either
of the two Eqs.~(\ref{gau57}) and ~(\ref{F285}) one obtains:
\begin{eqnarray}
\frac{1}{g_{YM}^{2} (\epsilon) } + \frac{N}{8\pi^2} \log
\frac{y^{2}}{\epsilon^2}  \equiv \frac{1}{g_{YM}^{2} (y ) }~~,~~
\epsilon^2 \equiv 2 \pi (\alpha' \Lambda )^2
\label{run2}
\end{eqnarray}
Eq.~(\ref{run2}) gives the one-loop correction to the bare gauge
coupling constant $g_{YM} (\Lambda )$ in the gauge theory
regularized with the cutoff $\Lambda$. By performing the
renormalization procedure one gets the renormalized coupling
constant:
\begin{eqnarray}
 \frac{1}{g_{YM}^{2} (\mu)}  - \frac{N}{8\pi^2} \log
\frac{\mu^{2}}{m^2} = \frac{1}{g_{YM}^{2} (m)}~~~;~~~
m^2 \equiv \frac{y^2}{2 \pi \alpha^{'2}} .
\label{run24}
\end{eqnarray}
from which  one can determine the
one-loop $\beta$-function
of ${\cal{N}}=2$ super Yang-Mills.

The vacuum angle $\theta_{YM}$ is provided by the terms in
Eqs.~(\ref{zetatet}) and ~(\ref{closed}) with the topological
charge. If we extract it from either of these two equations, we
find that it is imaginary and, moreover, must be renormalized like
the coupling constant. A way of eliminating these problems is to
introduce a complex cutoff and to consider both the configurations
in which the gauge field is in either one of the two sets of
branes. By taking the symmetric combination of these two
configurations
and introducing a complex cutoff $ \Lambda \rightarrow \Lambda
{e}^{-i
  \theta} $, the divergent
integral in Eq.~(\ref{gau57}), or equivalently in Eq.~(\ref{F285}), becomes:
\begin{eqnarray}
I (z) \equiv \int_{1/\Lambda^2}^{\infty}  \frac{d\sigma}{\sigma}
e^{-\frac{y^2 \sigma}{2\pi (\alpha')^2}} \simeq \log \frac{2 \pi
  (\alpha')^2 \Lambda^2}{y^2 {e}^{ 2i \theta}}~.
\label{comple68}
\end{eqnarray}
This procedure leaves unchanged all the previous considerations concerning
the gauge coupling constant, because in this case the coefficient of
the $F^2$ term results to be proportional to the
following combination:
\begin{eqnarray}
\frac{1}{2} \left[ I (z) + I ({\bar{z}}) \right] \simeq \log \frac{2 \pi
  (\alpha')^2 \Lambda^2}{y^{2}}~~.
\label{gau73}
\end{eqnarray}
In the case of the $\theta_{YM}$ angle one gets instead:
\begin{eqnarray}
\theta_{YM} = i\frac{N}{2} \left [ I (z) - I ({\bar{z}}) \right] =
2 N\theta~~~,
\label{theta45}
\end{eqnarray}
exactly reproducing the ${\cal N}=2$ SYM chiral anomaly.

\section
{{\bf{Branes in External Fields: ${\cal N} =1$ orbifold}}}
\label{sez4}

The analysis performed in the previous
Section has been extended
to the case of the orbifold $C^3/(  Z_2\times Z_2)$ 
preserving four supersymmetry charges \cite{DLMP}. The orbifold group
$ Z_2\times Z_2$ contains four elements whose action on
the three complex coordinates
\begin{eqnarray}
&&z_1=x_4+ix_5 \qquad z_2=x_6+ix_7 \qquad z_3=x_8+ i x_9
\label{z123}
\end{eqnarray}
is chosen to be:
\begin{eqnarray}
&&R_e= {\rm diag} \left(1,\,1,\,1\right) \qquad  R_{h_1}={\rm diag}\left(1,\,-1,\,-1\right)\nonumber\\
&& R_{h_2}= {\rm diag} \left(-1,\,1,\,-1\right) \qquad  R_{h_3}={\rm diag}\left(-1,\,-1,\,+1\right) .
\end{eqnarray}
The group
$ Z_2\times Z_2$ has four
irreducible
one-dimensional representations corresponding to four different
kinds of  fractional branes.
The interaction between a stack of $N_I$ ($I~=~1,\dots, 4$) branes of type $I$ and a D3-fractional brane, for
example of type $I=1$, with an $SU(N)$ gauge field turned on its
world-volume,
is the sum of four terms
\begin{eqnarray}
Z= Z_{e} + \sum_{i=1}^{3} Z_{h_i}~~,    \label{ZN=1}
\end{eqnarray}
where each contribution is obtained from the corresponding one in
the orbifold $ C^2/ Z_2 $ by adding an extra factor
$1/2$, due to the orbifold projection, and an index $i$ that labels
the elements of the orbifold group.

By extracting the coefficient of the term $F^2$ and performing on it the
field theory limit, one gets the correct running coupling constant and
chiral
anomalies
for ${\cal N}=1$ SYM, which are:
\begin{eqnarray}
&& \frac{1}{g_{YM}^{2} (\Lambda)} =
\frac{1}{g_{YM}^{2} (\mu) } + \frac{3N_1 - N_2 -N_3 -N_4}{16\pi^2} \log
\frac{\Lambda^{2}}{\mu^2}~~.
\label{run231}\\\
&&\theta_{YM}= \left(3N_1 - N_2 -N_3 -N_4\right)\theta. \,
\label{theta451}
\end{eqnarray}
showing again that the supergravity solution, dual to ${\cal N}=1$
SYM theory, gives the correct answer for the perturbative behavior
of the non conformal world volume theory.

\end{document}